\documentclass[twocolumn,prd,byrevtex,amsfonts,amssymb,amsmath,showpacs,letterpaper]{revtex4}

\usepackage{aas_macros}
\usepackage{times,helvet,varioref,color}
\usepackage{epsfig}
\usepackage{multirow}
\def \Msun{\ {\rm M_\odot}}
\def \neff{n_{\text{eff}}}
\def \ELL{\mathcal{L}}
\newcommand{\Eqref}[1]{Eq.~(\ref{#1})}
\newcommand{\Figref}[1]{Fig.~\ref{#1}}
  
\newcommand{\Tableref}[1]{Table~\ref{#1}}
\defcitealias{nfw}{NFW}

\begin{document}

\title{Can substructure in the Galactic Halo explain the ATIC and PAMELA results?}

\author{Pascal J.~Elahi$^1$, Lawrence M.~Widrow$^1$, Robert J.~Thacker$^2$}
\affiliation{$^1$Department of Physics, Engineering Physics \& Astronomy,
Queen's University, Kingston, Ontario, Canada; pelahi@astro.queensu.ca,
widrow@astro.queensu.ca\\
$^2$Department of Astronomy \& Physics, Saint Mary's University, Halifax, Nova
Scotia, Canada; thacker@ap.stmarys.ca}

\begin{abstract}
Recently, ATIC and PAMELA measured an anomalously large flux of leptonic cosmic rays which may arise from dark matter self-annihilation. While the annihilation signal predicted for a smooth halo is $10^2-10^3$ times smaller than the measured excess, the signal can be boosted by the presence of subhalos. We investigate the feasibility of large boost factors using a new Monte Carlo calculation technique that is constrained by previous simulation work on halo substructure. The model accounts for the observed decrease in the amount of substructure with decreasing halo mass and the scatter in halo structural parameters such as the density concentration parameter. Our results suggest that boost factors of $\sim10^2$ are ruled out at $\gtrsim14\sigma$. We conclude that substructure alone, at least with commonly assumed annihilation cross-sections, cannot explain the anomalous flux measured by ATIC and PAMELA.
\end{abstract}

\maketitle

%---------------------------------
Recent measurements by ATIC \cite{aticresults2008} and PAMELA \cite{pamelaresults2009} offer the tantalizing prospect that dark matter has been discovered, albeit indirectly. Both experiments have reported an anomalously large flux in leptonic cosmic rays at energies above $10$~GeV. Dark matter candidates, such as WIMPs (Weakly Interacting Massive Particles which arise in theories of supersymmetry) or Kaluza-Klein particles, can annihilate and produce cosmic rays \cite{bertone2005}. The excess flux might comprise secondary particles from annihilation events in the Galactic halo, a possibility that is now attracting considerable attention.

\par
Though the observed anomalous flux is 100-1000 times larger than the flux predicted from a smooth Milky Way halo with a standard thermal WIMP particle, the presence of subhalos can boost the signal since the annihilation rate is proportional to the square of the density. But while there is little doubt that dark matter halos are clumpy, the actual boost factor is a matter of some debate. In this paper, we take a critical look at the underlying assumptions in boost factor calculations and discuss the implications of our analysis for the ATIC and PAMELA results as well as current and future observations by the Fermi Gamma-Ray Large Area Space Telescope (GLAST) \cite{fermiresults2009,strigari2007,pieri2008}.

\par
Halos in cosmological N-body simulations host numerous subhalos \cite{moore1999}, which in turn host their own subhalos \cite{diemand2007}. The distribution of subhalos can be summarized by the mass function, $dN/d\ln f$, where $f\equiv m/M_h$ is the relative mass fraction of a subhalo of mass $m$ in a host of mass $M_h$. Note that in this paper, a host can refer to either a halo or a subhalo. The mass function appears to be well-characterized by a power-law \cite{gao2004,diemand2004},
\begin{equation}
    dN/d\ln f= Af^{-\alpha},\label{eqn:dndm}
\end{equation}
for $f<10^{-2}$. While current simulations probe $dN/d\ln f$ for $f\gtrsim10^{-6}$, a Galactic-mass halo with neutralino dark matter should have subhalos down to $f\sim10^{-18}$ \cite{martinez2009}, with the possibility of up to nine nested levels of substructure.
\begin{figure*}
    \centering
    \includegraphics[width=0.75\textwidth]{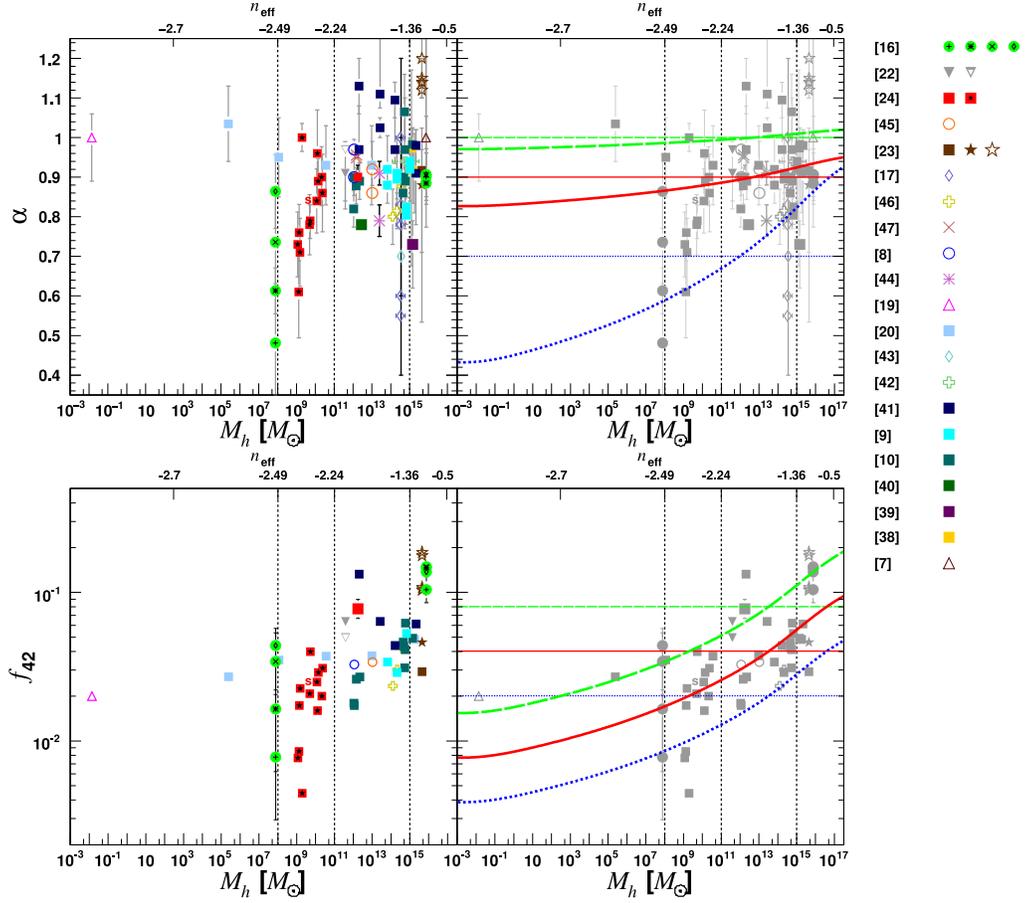}
    \caption{Compilation of $\alpha$ (top) and $f_{42}$ (bottom) from the studies indicated in right-hand key, where the year of the study decreases as one goes down the column. For both the upper and lower left panels, marker colors indicate different studies while the marker type indicate the algorithm used to identify subhalos. Studies which examined subhalo hosts instead of field halo hosts are indicated by internal black filled star. Colored horizontal error bars indicate mass range examined in study and black dashed vertical lines going from left to right indicate mass dwarf spheroidal galaxies, galaxies, and galaxy clusters. In the upper and lower right panels, we plot our phenomenological models of $\alpha$ and $f_{42}$ with the thick and thin lines for model 1 and 0 respectively. For $\alpha$, the mean, upper and lower values are given by the solid, dashed and dotted lines respectively. For $f_{42}$, the mean of the lognormal distribution is given by the solid line and $+2\sigma$ by the dashed, $-2\sigma$ by the dotted lines. Top panels: vertical black and grey error bars indicate uncertainty in the published and fit-by-eye values respectively. Bottom panels: black vertical error bars indicate variation between similar mass hosts.}
    \label{fig:subdata}
\end{figure*}

\par
The luminosity of annihilation secondaries (electrons, positrons, or photons) from a halo or subhalo of mass $M_h$ can be written in the form
\begin{equation}
    L_{e^\pm,\gamma}=\mathcal{P}_{e^\pm,\gamma}\frac{\langle\sigma
v\rangle}{m_\chi}\ELL(M_h),
\end{equation}
where $\mathcal{P}_{e^\pm,\gamma}$ are the branching ratios for electron-positron or photon secondaries, $\sigma$ is the total annihilation cross section, and $v$ and $m_\chi$ are the relative speed and mass of the annihilating particles. The quantity $\ELL(M_h)$ is the volume integral of $\rho^2$ where $\rho$ is the dark matter density. The boost factor due to subhalos is defined implicitly by the relation
\begin{equation}
    \ELL(M_h)=[1+B(M_h)]\tilde{\ELL}(M_h).\label{eqn:ell}
\end{equation}
Here $\tilde{\ELL}(M_h)$ represents the volume integral for a smooth halo, given by the volume integral of the smooth density field $\tilde{\rho}^2$. The boost factor is calculated recursively through the expression
\begin{equation}
    B(M_h)=\frac{1}{\tilde{\ELL}(M_h)}\int \frac{dN}{d\ln f}[1+B(m)]\tilde{\ELL}(m) d\ln f,\label{eqn:boost}
\end{equation}
where $m=fM_h$.

\par
Gamma-ray secondaries travel directly from the source to the observer and therefore the flux measured at Earth due to a source located at a distance $d$ is $L_\gamma/4\pi d^2$. In contrast, charged particles are scattered by the Galactic magnetic field and therefore the arrival direction of the secondaries is essentially independent of the direction to the source. Furthermore, charged particles lose energy en route from the source to the observer. It is therefore useful to define an effective (energy-dependent) total boost due to all subhalos (see Ref.~\cite{lavalle2008b} for details):
\begin{align}
    B_{\text{eff}}&(E;M_h,E_i,{\bf x}_\odot)=(1-f_t)^2\notag\\
        &+\frac{\mathcal{G}_{\rm sub}(E;E_i,{\bf x}_\odot)\xi_{\rm sub}(M_h)}{\xi_{\rm halo}(E;E_i,{\bf x}_\odot)}.\label{eqn:beff}
\end{align}
Here $f_t$ is the total mass fraction in subhalos and again $m=fM_h$. The first term accounts for the fact that $\tilde\rho$ is defined such that $\int\tilde{\rho}dV=M_h$, that is the halo is smooth, but only $(1-f_t)$ of the host's mass is in the smooth radial component. Essentially $(1-f_t)^2$ is a normalization term.

\par
In \Eqref{eqn:beff}, the effect of subhalos has been broken up into two terms by assuming that the volume distribution of subhalos is independent of mass distribution. The first term in the numerator, $\mathcal{G}_{\rm sub}$, accounts for the propagation of cosmic rays originating from subhalos and is given by
\begin{align}
    \mathcal{G}_{\rm sub}(E;E_i,{\bf x}_\odot)&=\int G(E,{\bf x}_\odot;E_i,{\bf x}_i)\frac{\tilde{\rho}({\bf x}_i)}{M_h}d^3{\bf x}_i\label{eqn:gsub},
\end{align}
where the host term has been normalized by the solar density $\rho_\odot$. Here the propagation of the secondaries from an initial position ${\bf x}_i$ with initial energy $E_i$ to the Earth with final energy $E$ is explicitly accounted for using the Green function $G(E,{\bf x}_\odot,E_i,{\bf x}_i)$. The propagation is a diffusive process and consequently $G(E,{\bf x}_\odot,E_i,{\bf x}_i)\propto\exp\left[-({\bf x}_i-{\bf x}_\odot)^2/\lambda_D^2(E;E_i)\right]$ where $\lambda_D$ is the diffusion length. For electron-positron pairs produced by a $100-1000$~GeV WIMP, $\lambda_D\sim$~few kpc for $E\gtrsim100~$GeV and decreases monotonically as $E\rightarrow E_i$ (see Ref.~\cite{lavalle2008b} for further discussion). In the definition of $\mathcal{G}_{\rm sub}$, we are effectively treating subhalos as point sources, which is a reasonable assumption as $\approx90\%$ of the flux originates from the central region of a (sub)halo. As we are interested in subhalos that are numerous enough to enhance the diffusive background, we are generally concerned with subhalos with masses of $\lesssim10^8\Msun$. These halos have central regions that are $\lesssim$~kpc in radius. We also assume in \Eqref{eqn:gsub} that subhalos trace the host's dark matter distribution.

\par
The second term in the numerator, $\xi_{\rm sub}$, is the total contribution of subhalos to the annihilation flux and is given by
\begin{align}
    \xi_{\rm sub}(M_h)=\int \frac{dN}{d\ln f}\bar{\ELL}(m)[1+B(m)] d\ln f.\label{eqn:btot}
\end{align}
This is equivalent to the cosmic ray source term. The main difference between our work and previous studies \cite{lavalle2008b,pato2009,kuhlen2009a} is that we examine the enhancement that arises explicitly due to the entire subhalo hierarchy, that is subhalos, subsubhalos, etc., by incorporating the boost factor.

\par
The $\xi_{\rm halo}$ term in denominator of \Eqref{eqn:beff} is the host's contribution to the cosmic ray flux which also accounts for the propagation of cosmic rays and is given by 
\begin{align}
    \xi_{\rm halo}(E;E_i,{\bf x}_\odot)&=\int G(E,{\bf x}_\odot;E_i,{\bf x}_i) \tilde{\rho}^2({\bf x}_i)  d^3{\bf x}_i\label{eqn:ghost},
\end{align}
Here again the propagation of the secondaries is explicitly accounted for using the Green function $G(E,{\bf x}_\odot,E_i,{\bf x}_i)$.

\par
The effective boost in cosmic ray flux can be greatly enhanced by the boost factor. The boost factor depends sensitively on the subhalo mass function. Using the often quoted values $\alpha=1$ and $A=0.033$ \cite{kuhlen2008}, one finds $B\approx 30$ for $\gamma$-rays with a Galactic-mass halo. The key assumption in obtaining this result is that $\alpha$ and $A$ are independent of the host mass and apply to all scales and levels in the subhalo hierarchy. Our goal is to examine the validity of this assumption, in short, to test whether $\alpha$ and $A$ depend on the mass of the host. To do so, we compare estimates for $\alpha$ and $A$ over a wide range of published simulations.

\par
Though most of the results in our study are based on simulations which assume a standard $\Lambda$CDM cosmology, results from our own simulations of structure formation in scale-free cosmologies \cite{elahi2009} as well as results from simulations of warm dark matter cosmologies \cite{knebe2008} are also included. To assign an effective mass to hosts in these non-standard simulations we appeal to the halo formation process. In the hierarchical clustering paradigm, halos arise from primordial density fluctuations where small-scale structures form first and then merge together to form larger and larger objects. This process is governed by the power spectrum of the perturbations, $P(k)$, where $k$ is the comoving wavenumber of a perturbation mode. Alternatively, one can describe the perturbation spectrum by the mass variance, 
\begin{align}
    \sigma^2(M)=\int \frac{k^2dk}{(2\pi)^3}P(k)W^2(k,M),
\end{align}
where $W(k,M)$ is the window function enclosing a mass $M\propto k^{-3}$ within a comoving volume of radius $r=2\pi/k$. This quantity is simply the average rms overdensity of a sphere enclosing a mass $M$.

\par
In scale-free cosmological models the primordial power spectrum is a power-law function of wavenumber $k$, that is, $P\propto k^n$ where $n$ is referred to as the spectral index. This form of the power spectrum leads to a mass variance obeying $d\ln\sigma^2/d\ln M = -(n + 3)/3$. Though the power-spectrum in a $\Lambda$CDM cosmology is more complicated, we can define an effective spectral index, $\neff(M)\equiv-3(d\ln\sigma^2/d\ln M+1)$. Thus, the spectral index in a scale-free cosmology may be used as a proxy for the ($\Lambda$CDM) mass scale \cite{elahi2009} by setting $n = \neff(M)$ and solving for $M$. In this work we take $\Omega_m=0.25$, $\Omega_\Lambda=0.75$, $h=0.73$, $\sigma_8=0.9$ and $n_s=1$ as our reference cosmology.

\par
The use of different algorithms to identify subhalos introduces a further complication.  These algorithms range from SUBFIND \cite{subfind}, which associates subhalos with local density peaks, to 6DFOF \cite{diemand2006,elahi2009}, which searches for clustering in 6D~phase space. In addition, some (but not all) researchers apply an unbinding criterion which removes unbound particles when estimating the mass of a subhalo. One reason for such a variety of methods is that subhalo identification is an ill-defined problem.

\par
To characterize the amount of substructure, we introduce $f_{42}$ in place of $A$, where 
\begin{equation}
    f_{42}=\int\limits_{-4\ln10}^{-2\ln10} f\frac{dN}{d\ln f} d\ln f,
\end{equation}
is the fraction of mass in subhalos with $10^{-2}<f<10^{-4}$. In \Figref{fig:subdata} we plot estimates of $\alpha$ and $f_{42}$ from a large sample of studies. When available, we plot published values of $\alpha$. Otherwise, we fit-by-eye for the mean $\alpha$ and estimate the uncertainty. The mean $\alpha$ is then used to determine $f_{42}$. Error bars for $f_{42}$ are shown when a study reports the variation in the subhalo number or mass fraction across several hosts of the same mass.

\par
The figure reveals a large amount of scatter in $\alpha$ and $f_{42}$. For galactic to cluster masses, estimates of $\alpha$ range from $0.7$ and $1.1$, which is generally within the estimated uncertainty in $\alpha$ for a single measurement and the variation in $\alpha$ with redshift for an individual halo \cite{gao2005,elahi2009}. The scatter in $\log f_{42}$ is roughly constant with $M_h$ with most points lying within $0.15$~dex of the mean. The scatter may be due to variations in the mass accretion histories of the halos. The halo's mass accretion history is important not only for defining the density parameter \cite{tasitsiomi2004} but it is also responsible for delivering new substructure within a halo.

\par
Figure \ref{fig:subdata} also suggests that $\alpha$ and $f_{42}$ decrease with decreasing $M_h$. The observed $M_h$-dependence should be treated with some caution. For example, a linear fit to $\alpha(M_h)$ for $M_h$ between galaxy and cluster masses is consistent at the $1\sigma$-level with $\alpha=$~constant. Furthermore, at smaller host halo masses ($\neff$ closer to $-3$) the subhalo mass function is more sensitive to the type of subhalo-finding algorithm used \cite{elahi2009} and in particular, whether a binding criterion is applied. For smaller $M_h$, subhalos tend to be less gravitationally bound. Studies that do not correct for unbound particles, such as \cite{diemand2006}, may overestimate subhalo masses and therefore overestimate $\alpha$ and $f_{42}$. Even for larger $M_h$, the application of a correction for unbound particles decreases $\alpha$ by $0.1-0.2$ (see, for example, Ref.~\cite{athanassoula2009,maciejewski2008}). It should also be noted that the data points from Ref.~\cite{springel2008} at subgalactic masses are for subhalo hosts. This study found that the mass fraction of subsubhalos in subhalos tends to be smaller than the mass fraction of subhalos in halos. It may well be that $\alpha$ depends on the host's level in the subhalo hierarchy, that is $\alpha$ of a field halo may differ from that of a similar mass host that sits within the virial radius of a larger halo.

\par
Most boost factor calculations assume $\alpha$ and $f_{42}$ are independent of mass, that is, $\alpha(M_h)=\alpha_{o}$, $f_{42}(M_h)=f_o$. However, this assumption does not capture the trends in \Figref{fig:subdata}, especially at lower $M_h$, and we therefore propose the following phenomenological model:
\begin{align}
    \alpha(M_h)&=\alpha_o+\alpha_n\log[\neff(M_h)+3], \label{enq:amod1}\\
    \log f_{42}(M_h)&=\log f_o+\log f_n\log[\neff(M_h)+3].\label{eqn:fmod1}
\end{align}
We account for the scatter in $\alpha$ by considering three values for each $\alpha_o$ and $\alpha_n$ - the mean and the upper and lower envelop values as given in \Tableref{tab:param}. For $f_{42}$ we fit the mean by eye and assume that it follows a lognormal distribution. This choice is motivated by the fact that other bulk halo parameters, such as the concentration parameter, follow lognormal distributions \cite{maccio2008}.

\begin{table}
\centering
\caption{Summary of ($\alpha,f_{42}$) model parameters}
\begin{tabular*}{0.45\textwidth}{@{\extracolsep{\fill}}cccccc}
\hline
\hline
    Model \#  & \Figref{fig:boost} color & $\alpha_o$ &$\alpha_n$ & $f_o$ & $f_n$ \\
\hline
    0-u & g  & $1.0$  & $0$   & $0.04$ & $0$\\
    0   & r  & $0.9$  & $0$   & $0.04$ & $0$\\
    0-l & b  & $0.7$  & $0$   & $0.04$ & $0$\\
    1-u & g  & $1.0$  & $0.046$ & $0.034$  & $10$\\
    1   & r  & $0.9$  & $0.12$ & $0.034$  & $10$\\
    1-l & b  & $0.7$  & $0.46$ & $0.034$  & $10$
\end{tabular*}
\label{tab:param}
\end{table}

\par
We calculate the boost factor using \Eqref{eqn:boost}, under the assumption that the subhalo mass function is a power-law with an index $\alpha(M_h)$. The amplitude is normalized using $f_{42}(M_h)$ and $\alpha(M_h)$ subject to the condition that the total fraction of mass in subhalos is less than one. The mass function extends down to a mass $m_o$, the minimum CDM halo mass, which depends on the properties of the dark matter particle. Typical values for $m_o$ when dark matter is a neutralino in the Constrained Minimal Supersymmetric Standard Model are $10^{-9}-10^{-6}\Msun$ \citep{martinez2009}, though the minimum CDM halo mass can vary between $10^{-12}-10^{-4}\Msun$ \cite{profumo2006}. We use the more probable value of $m_o=10^{-6}\Msun$ \citep{martinez2009}. We also assume that a host cannot contain subhalos with $f>10^{-2}$ as suggested by most studies. As a consequence, the smallest host mass that contains subhalos is $100m_o$. Finally, we only go to three levels in the subhalo hierarchy since deeper levels change the boost factor by $\lesssim1\%$. As a general observation the mass flux tends to come from the first two levels of the subhalo hierarchy.

\par
Due to the condition that $f_t\leq1$, the boost factors begin to saturate for large $f_{42}$ and small $m_o$. The limiting values of $f_{42}$ for a given $\alpha$ and $m_o/M_h$ is given in \Tableref{tab:f42lim}. This table shows that the limits are only important for $\alpha>1$ and only when $m_o/M_h$ is large, for example, when $m_o=10^{-12}\Msun$ for $M_h=10^{12}\Msun$. For our choice of $m_o=10^{-6}\Msun$, galactic mass hosts, and our subhalo mass function models, this saturation occurs at a $f_{42}$ that is still $\approx3\sigma$ away from the mean even at $\alpha=1.0$.

\par
The volume integral of $\tilde\rho^2$ in \Eqref{eqn:ell} is given by
\begin{equation}
    \tilde{\ELL}(M_h)=\frac{\rho_v}{3}M_h c^3(M_h) g(c),\label{eqn:ellbar}
\end{equation}
where $\rho_v$ is the virial density of a halo, $c\equiv r_v/r_s$ is the concentration parameter, $r_s$ is the profile's effective radius, and the dimensionless function $g(c)$ depends on the form of the density profile. We consider the commonly used \citetalias{nfw} \cite{nfw} profile, which is described by two parameters, the characteristic radius $r_s$ and density $\rho_s$. We also examine the Einasto profile, which appears to be a better fit to halos from cosmological simulations \cite{springel2008}. The Einasto profile has an extra parameter describing the radial logarithmic slope of the density profile which appears to have a mass dependence \cite{gao2008}. We find that an Einasto profile increases the boost factor for subgalactic masses by $\lesssim50\%$. To determine the concentration parameter we use the two-parameter model presented in Ref.~\cite{maccio2008} for $c(M)$ as this model provides a better match to simulation data than the often-used model of Ref.~\cite{bullock2001}. Using the Ref.~\cite{maccio2008} model for $c(M)$ instead of the Ref.~\cite{bullock2001} model reduces $B$ by $\lesssim10\%$. 
\begin{table}
\centering
\caption{Limits on $f_{4    2}$ due to $f_t\leq1$}
\begin{tabular*}{0.325\textwidth}{@{\extracolsep{\fill}}ccc}
\hline
\hline
    $\alpha$ & $m_o/M_h$ & $f_{42,{\rm lim}}$ \\
\hline
    0.8 & $10^{-24}$ to $10^{-16}$ & 0.601 to 0.602 \\
    0.9 & $10^{-24}$ to $10^{-16}$ & 0.371 to 0.384\\
    1.0 & $10^{-24}$ to $10^{-16}$ & 0.091 to 0.143\\
\end{tabular*}
\label{tab:f42lim}
\end{table}

\par
We calculate $10^4$ random realizations of $B$ for a given $M_h$ where $c$ and $f_{42}$ are sampled from lognormal distributions with $\sigma_{\log f_{42}}=0.15$~dex and $\sigma_{\log c}=0.10$~dex \cite{neto2007}. Due to the recursive nature of \Eqref{eqn:boost}, variation in $c$ not only affects $\bar\ELL$ of the host but that of the host's subhalos. The distribution of $B(M_h)$ across the $10^4$ realizations appears to be well characterized by a lognormal distribution. We fit this distribution for the mean and the variance.

\par
\begin{figure*}
    \centering
    \includegraphics[width=0.995\textwidth]{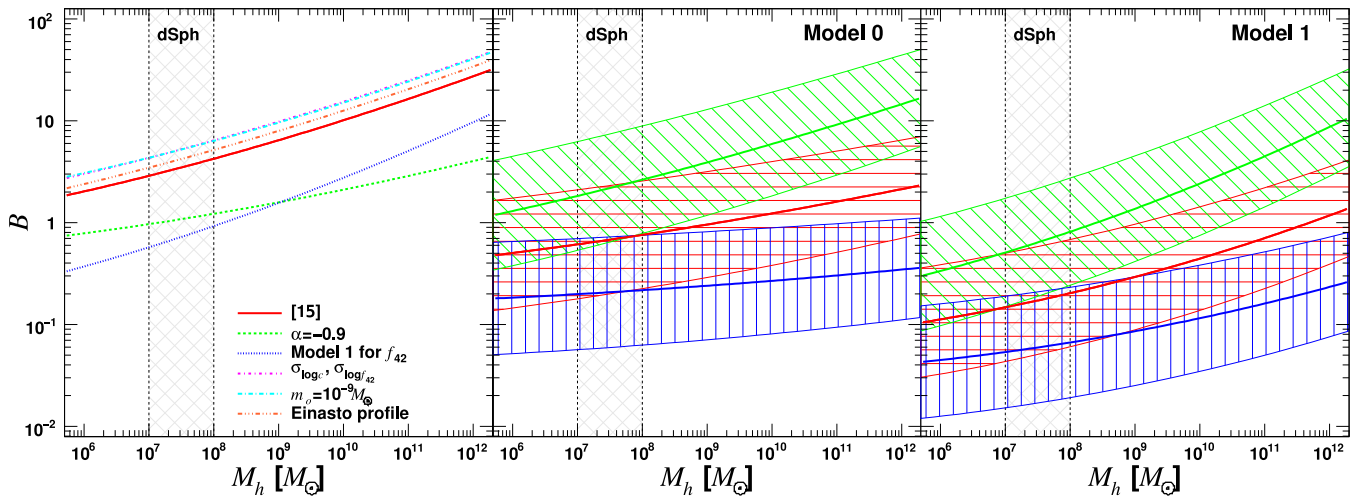}
    \caption{Boost factors. Left panel shows how $B(M_h)$ changes when a single parameter is changed. Middle (right) panel shows model 0 (1) with thick red, green and blue lines indicating mean boost factor using the mean, upper, and lower values of $\alpha$ respectively. Colored hatched contours indicate the $2\sigma$ region arising from the variation in $\log c$ \& $\log f_{42}$. The gray hatched region outlines the mass range for dwarf spheroidal satellite galaxies of MW.}
    \label{fig:boost}
\end{figure*}

The results are plotted in \Figref{fig:boost}. We particularly highlight the mass scales of the Galaxy's dwarf spheroidal satellites as these objects may be the best candidates for searches of $\gamma$-rays secondaries \cite{strigari2007}. The left panel shows how a few key parameters individually affect $B(M_h)$. As a reference model, we use $m_o=10^{-6}\Msun$, $\alpha=1.0$, $f_{42}=0.067$, and an \citetalias{nfw} density profile with the Bullock prescription for $c(M)$ \cite{kuhlen2008}. Note that in Ref.~\cite{kuhlen2008}, the boost factor is calculated for subhalos below the mass resolution of their {\it Via Lactea II} simulation, which is $\sim10^5\Msun$. This amounts to calculating \Eqref{eqn:boost} with the upper limit given by $10^{5}\Msun/M_h$. Decreasing $\alpha$ from $1.0$ to $0.9$, or using the mean values from our new phenomenological model for $f_{42}$ decreases $B(M_{\text{dSph}})$ by $\sim4$. The introduction of scatter in $c$ and $f_{42}$ increases the mean boost factor slightly, though the amount depends on the exact form and width of the distributions. For our choices, $B(M_{\text{dSph}})$ increases by $\approx30\%$. Decreasing $m_o$ to $10^{-9}\Msun$ also increases the boost factor by $\approx30\%$ while using an Einasto profile increases the boost factor by $\lesssim20\%$ in this case.

\par
The other two panels of \Figref{fig:boost} show the peak of the distribution and the $2\sigma$ contours from the models listed in \Tableref{tab:param}. We find increasing $\alpha$ increases the boost factor by $\approx2$ for dSph galaxies, though the differences are generally within the variation caused by $c$ and $f_{42}$. For model 0, the mean $\alpha$ gives $B(M_{\text{dSph}})\sim0.6^{+1.4}_{-0.4}$, though for $\alpha=1.0$, $B\sim8$ is within the $2\sigma$ envelop. Model 1 reduces the boost factors such that $B(M_{\text{dSph}})\gtrsim2$ lie outside the $2\sigma$ envelop with the mean $\alpha$ giving $B(M_{\text{dSph}})\approx0.2$.

\par
Thus, subhalos in the Galaxy's satellites are unlikely to greatly enhance the $\gamma$-ray flux. Previous estimates of the number of satellites that could be detected via their $\gamma$-ray flux by GLAST, such as those by Ref.~\cite{strigari2007,kuhlen2008}, are probably overly optimistic. Even our revised calculations may overestimate the boost factor as it appears that subhalos have smaller $f_{42}$ than similar mass halos. Consequently, satellites might even have their boost factors reduced by a factor of $\gtrsim2$. So far, GLAST has detected numerous $\gamma$-ray sources, none of which are convincing dark matter annihilation signals \cite{fermigrlist2009}.

\par
We now examine the consequences of such boost factors for cosmic rays. To determine the energy dependence in \Eqref{eqn:beff}, we use the mean propagation parameters listed in Ref.~\cite{lavalle2008b} and assume that the dark matter particle annihilates directly to monoenergetic $e^{\pm}$. For a $700$~GeV thermal dark matter particle with $\langle\sigma v \rangle=3\times10^{-26}~\text{cm}^3\text{s}^{-1}$, the diffusion length is $1.3$~kpc at $E=300$~GeV and monotonically decreases to $0.1$~kpc at $E=690$~GeV. We also assume that the Milky Way halo has an \citetalias{nfw} density profile with a characteristic radius $r_s=20$~kpc and a solar density of $\rho_\odot=0.43$~GeV cm$^{-3}$.

\par
Recall that in \Eqref{eqn:beff} we assumed that the subhalo volume distribution is the same as the halo's density profile. However, simulations appear to show that subhalos do not trace the host halo's radial density profile \cite{springel2008,diemand2009}. Instead tidal stripping destroys subhalos in the center, resulting in a radially anti-biased distribution, at least for subhalos with masses of $\gtrsim10^5\Msun$. Very few subhalos are thus found within $\sim20$~kpc of the halo center. This implies is that subhalos are unlikely to contribute to the observed flux as they are too far away. We argue that it is unreasonable to assume that this anti-biased radial distribution continues down to very small masses since these subhalos should become increasingly less susceptible to tidal disruption as their mass decreases and therefore should be able to survive to smaller radii. It is also worth noting that it is difficult to identify substructure in the central, high density regions of halos with current group finder algorithms. Furthermore, Ref.~\cite{diemand2009} found that though tidal fields affect the mass of subhalos, their annihilation luminosity is less affected and more closely follows the host halo's radial density profile. Thus, we compromise and only include contributions from subhalos with $m\lesssim10^4\Msun$. Considering only low mass subhalo hosts reduces the contribution of substructure by $\approx60\%$ for both models 0 and 1. This may still be an optimistic assumption since interactions with baryonic matter, that is the Galactic disk and stars, might further reduce the inner subhalo abundance.

\par
In general we find $B_{\text{eff}}(E\geq300\text{~GeV})\sim1$. Even the most optimistic model, 0-u, gives $B_{\text{eff}}(E\geq300\text{~GeV})=1.3^{+0.6}_{-0.4}$. Here the variation in $B_{\text{eff}}$ is due primarily to the variation in $f_{42}$ for the Galactic halo based on our phenomenological model, though we also account for the variation in the boost factors and concentration parameters of subhalos. This model excludes the $B_{\text{eff}}\sim200$ required to explain the ATIC signal at the $\sim14\sigma$ level. These results are in agreement with similar studies \cite{lavalle2008b,pato2009,kuhlen2009a}. We have improved on earlier results by explicitly including several levels in the subhalo hierarchy that earlier work neglected. 

\par
The High Energy Stereoscopic System (HESS) \citep{hessresults2009} and Fermi GLAST \citep{fermiresults2009} found different though not necessarily contradictory results. Both instruments detect an excess, though not as large as that observed by ATIC, and neither instrument observes the pronounced peak at $\sim700$~GeV seen by ATIC. The Fermi GLAST measurements require a boost of $\sim150$ at $E=300$ which monotonically decreases to $\approx70$ at $E=600$. This is still rejected by our model at the $10-12\sigma$ level.

\par
Some studies have suggested that it is possible to reproduce the observed flux, particularly the ATIC bump in the energy spectrum, with a single nearby subhalo \cite{hooper2009,kuhlen2009a}. The required annihilation rate is $\sim10^{37}$~s$^{-1}$ for a thermal WIMP with $m_\chi\sim100-1000$~GeV. Neglecting the boost factor from deeper levels in the subhalo hierarchy, a subhalo with a mass of $\gtrsim10^8\Msun$ at $\sim1$~kpc is required. However, Ref.~\cite{brun2009} find based on a numerical simulation that the probability of such a large subhalo within $\sim8$~kpc of the Galactic center is exceedingly low, $p\lesssim 10^{-5}$. This clearly appears to be highly unlikely. Incorporating the boost factors, and thereby incorporating the entire subhalo hierarchy, does not drastically reduce the minimum mass required to explain the ATIC peak and thereby the probability of a nearby clump.

\par
The anomalous cosmic ray flux observed by ATIC and PAMELA have sparked a flurry of interest as there is the possibility that these instruments may have indirectly detected annihilating dark matter. However, the observed amplitude cannot be explain with standard thermal dark matter. A number of studies have invoked subhalos to explain the large amplitude of this flux. Our work effectively ends this line of thought. We find that subhalos alone are unlikely to account for the anomaly and are strongly ruled out at the $14\sigma$ level. This still leaves the possibility that the flux is due to instrument errors, astrophysical phenomena, or, more intriguingly, perhaps exotic theories of dark matter are required \cite{arkani-hamed2009,kuhlen2009b}.

\acknowledgments
The authors thank the anonymous referees for useful comments. PJE acknowledges funding from the NSERC. RJT and LMW acknowledge funding by respective Discovery Grants from NSERC. RJT is also supported by grants from the Canada Foundation for Innovation and the Canada Research Chairs Program.

%-------------------------
\bibliography{boost.bbl}

\end{document}